\documentclass[twoside,twocolumn,9pt]{article}
\usepackage{graphicx} 
\usepackage{extsizes}
\usepackage[super,sort&compress,comma]{natbib} 
\usepackage[version=3]{mhchem}
\usepackage[left=1.5cm, right=1.5cm, top=1.785cm, bottom=2.0cm]{geometry}
\usepackage{balance}
\usepackage{times,mathptmx}
\usepackage{sectsty}
\usepackage{graphicx} 
\usepackage{lastpage}
\usepackage{ulem}
\usepackage[format=plain,justification=justified,singlelinecheck=false,font={stretch=1.125,small,sf},labelfont=bf,labelsep=space]{caption}
\usepackage{float}
\usepackage{fancyhdr}
\usepackage{fnpos}
\usepackage[english]{babel}
\addto{\captionsenglish}{%
  
}
\usepackage{array}
\usepackage{droidsans}
\usepackage{charter}
\usepackage[T1]{fontenc}
\usepackage[usenames,dvipsnames]{xcolor}
\usepackage{setspace}
\usepackage[compact]{titlesec}
\usepackage{hyperref}

\graphicspath{ {images/} }

\title{Multi-component colloidal gels: interplay between structure and mechanical proprieties }
\author{Claudia Ferreiro-C\'ordova,\textit{$^{a}$} Mehdi Bouzid,\textit{$^{b}$} Emanuela Del Gado,\textit{$^{c}$} and Giuseppe Foffi$^{\ast}$\textit{$^{a}$}}

\date{\today}
\begin{document}
\maketitle

\begin{abstract}
We present a detailed numerical study of multi-component colloidal gels interacting sterically and obtained by arrested phase separation. Under deformation, we found that the interplay between the different intertwined networks is key. Increasing the number of component leads to softer solids  that can accomodate progressively larger strain before yielding.
The simulations highlight how this is the direct consequence of the purely repulsive interactions between the different components, which end up enhancing the linear response of the material.  
Our work {provides new insight into mechanisms at play for controlling the material properties and open the road to new design principles for} soft composite solids
\end{abstract}

\section{Introduction}
\footnotetext{\textit{$^{a}$~Laboratoire de Physique des Solides, CNRS, Universit\'e Paris-Sud, Universit\'e Paris-Saclay, 91405 Orsay, France; E-mail: giuseppe.foffi@u-psud.fr}}
\footnotetext{\textit{$^{b}$~Laboratoire de Physique Th\'eorique et Mod\`eles Statistiques, CNRS, Universit\'e Paris-Sud, Universit\'e Paris-Saclay, 91405 Orsay, France.}}
\footnotetext{\textit{$^{c}$~Department of Physics, Institute for Soft Matter Synthesis and Metrology,
Georgetown University, Washington, D.C. 20057, USA.}}

Gels are ubiquitous in soft matter with applications that range from food~\cite{Mezzenga2005} to materials and biomaterials science~\cite{Peppas2006}. In these systems, the main feature is the presence of a ramified backbone that confers the capability to sustain finite stresses. The building blocks of the backbone (e.g. polymers~\cite{Osada1998}, proteins~\cite{Storm2005, gibaud2012}, colloids~\cite{Zaccarelli2007} or micelles) and their geometrical nature can alter the mechanical response and, for this reason, their design is of great technological importance. 
The advances in syntesis have allowed the fabrication of colloidal particles in a variety of sizes, shapes and functionality~\cite{Zaccarelli2007, Lu2013, Glotzer2007}, contributing to the increasing interest in the field.
Colloidal gels have been used for a wide range of applications such as synthetic colloid porous materials~\cite{duguet2011,loverso2006}, functionalisation of surfaces and films production~\cite{miljanic2008,wangq2008}, ceramics processing~\cite{wyss2005,schenker2008},  foams~\cite{Muth2017}, protein systems~\cite{cardinaux2007,vanGruijthuijsen2012}, food science~\cite{dickinson1992,Mezzenga2005}, soft matter~\cite{Zaccarelli2007,dorsaz2011} and bio-engineering~\cite{Nair2019}. {One common route to colloidal gelation is through arrested phase separation\cite{foffi2005,lu2008,Testard2011,gibaud2012,zia1,zia2,helgeson2015}}. In this scenario, attractive forces between spherical colloids, as originated by depletion or charge destabilisation, induce a thermodynamics phase separation between two liquids at different densities. The final equilibrium state, however, is never reached due to the onset of a kinetic arrest that ultimately leads  to an arrested space spanning percolating network: the colloidal gel \cite{bookchaptersuliana}.\\
Different models have been proposed for colloidal gels, including short range isotropic interactions\cite{lu2008,foffi2005,foffi2005JCP,zia1,zia2}, anisotropic effective interactions \cite{delgado-kob,colombo2014SM}, patchy models \cite{Seiferling2016,Rovigatti2011}, short-ranged non-central forces \cite{nguyen2019computer} and dipolar particles \cite{Blaak2007,ilg-dipolar,Goyal2010}.  The key idea  is to develop relatively simple, but microscopically relevant, particle models that can induce macroscopic structural disorder characterised by complex spatio-temporal fluctuations and correlations typical of kinetically arrested systems \cite{bouzid-bookchapter}. 

In the last decade, several studies have focused on understanding how the kinetics drives the self-assembly towards an out-of-equilibrium gel phase, which is now well understood thermodynamically~\cite{Zaccarelli2007, Lu2013, bookchaptersuliana}. However, the rheological properties of colloidal gels are still under active investigation~\cite{koumakis2011two,colombo2014,chirp2018,Bouzid2017,mao2019,Tsurusawa2019,zia-petekidis,swan}.   Typically, the non-linear mechanical response of gels is visco-elasto-plastic and  the prevalence of one behavior or the other may depend on the mode of deformation compared to the observation time. The result is a complex rheological behavior including shear banding, strain hardening, creep, fracture, slip, ageing, rejuvenation and yielding~\cite{Calzolari,Joshi2014,bouzid-bookchapter}. Indeed, when the shear stress is quite important, the system ends up either fluidising like yield stress fluids, or fracturing like soft solids (see for example also protein gels or (bio) -polymers) \cite{laurati, perge,divoux,gibaud,aime,cipelletti2019microscopic}. Thus, it is evident that understanding the physical mechanisms that control colloidal gel response to external stimuli is key to tune their macroscopic properties. 

The examples discussed so far are characterised by a single type of particle. However, it has been shown that increasing the number of components can change dramatically the {  morphology and connectivity} of gels. For binary mixtures, for example, by carefully tuning the inter-particle attractions it is possible to obtain double colloidal gels, known as bigels, that are the result of an arrested demixing \cite{Varrato2012}.  As a matter of fact, double network gel formation has been investigated in different systems, both experimentally and with simulations. These systems include patchy particles \cite{Seiferling2016,Rovigatti2011}, mixtures of dipolar particles \cite{Blaak2007,Goyal2010,ilg-dipolar}, fumed silica-based organogels \cite{Patel2015}, protein mixtures \cite{Blumlein2015} and selective short range interactions \cite{DiMichele2013,Varrato2012,DiMichele2014}. In all of them, interpenetrating networks of two different gels that seem to exhibit non-intuitive mechanical properties are formed\cite{DiMichele2013,Varrato2012,DiMichele2014,Blumlein2015,Patel2015}. In particular, aqueous-organic bigels and two-protein bigels have been reported to exhibit an elastic modulus higher than the sum of the elastic moduli of each of the two forming gels\cite{Blumlein2015,Patel2015}.\\

In the present work we use selective short-range interactions to model multiple component gels. The case of two component mixtures has already been studied and found to form bigels giving preliminary results of increase in the yield stress when compared with monogels\cite{DiMichele2013,Varrato2012,DiMichele2014}. {  Indeed double networks have been introduced as a strategy to obtain stronger and tougher materials in the context of hydrogels or elastomers \cite{Gong2010,creton} and hence those results suggest that similar strategies could be attempted for colloidal gels. Here, using a model for colloidal gels similar to \cite{Varrato2012}, we explore mixtures of up to three components to disentangle the impact of the structure, the dynamics and the mechanical response in these soft solids. In particular, we find that when the mutual interactions between the different particle species are prevalently repulsive, the multi-component gels are softer and allow for accommodating much larger strains, leading to potentially tougher but definitely softer materials.}

\section{Model and numerical simulations}\label{sec:model}

\subsection{Numerical model} 
 
We have developed a minimal 3D model for multi-component colloidal particle gels formed by a mixture of $N$ spherical particles, with $m$ distinguishable species of diameter $d$. Our model uses selective interactions: intra-species interactions $u_{ss}$ are given by a short range attractive well combined with a repulsive core \`a la Lennard-Jones widely used to study gelation\cite{lu2008,foffi2005} while inter-species interactions $u_{ss'}$ are set as purely repulsive. These interactions are described by the potential:

\begin{equation} \label{eq:potentiallq}
u_{s(i) s(j)}(r_{ij})= \left\{
  \begin{array}{l l}
     C \epsilon \left[ \left( \frac{d}{r_{ij}} \right)^{p}-\left( \frac{d}{r_{ij}} \right)^{q} \right]+ (1-\delta_{ss'})U_{0} & \quad r_{ij} < r^{ss'}\\\\
     0 & \quad r_{ij} \geq r^{ss'}
  \end{array}. \right.
\end{equation}

The position vectors of all particles are given by $\mathbf{\{r_i\}}$, $r_{ij} \!~=~\! |\mathbf{r_{j}} - \mathbf{r_{i}}|$ denotes the distance between two particles ($i$ and $j$) and $s(i)$ is the species of particle $i$. The additive term containing $U_0$ guarantees that the potential is zero at the cut-off.  All the particles have the same diameter $d\!~=~\!1$ which defines the length units while the energy units are defined such that $\epsilon\!~=~\!1$ for all species.

The exponents $p$ and $q$ in Eq.~\ref{eq:potentiallq} define the range of the attraction, and are set to $p\!~=~\!36$ and $q\!~=~\!24$. The constant $C\!~=~\!\frac{p}{p-q}\left(\frac{p}{q}\right)^{q/(p-q)}$ ensures that the minimum is at $\epsilon\!~=~\!1$ while the cut-off parameter $r^{ss'}$ sets the range of the interaction.  The cut off is chosen such that
\begin{equation} \label{eq:potentialcut}
r^{s s'}= \left\{
  \begin{array}{l l l}
    1 .6d &  s=s'& {\rm intra-species}\\\\
     1.03d& s\neq s' & {\rm inter-species}
  \end{array}. \right.
\end{equation}

For $s\neq s'$, the cut off distance corresponds to the minimum of the potential. In this way, the attractive well is present for same species while the potential is purely repulsive for different ones. The shift value $U_{0}$ is chosen accordingly such that $u_{ss'}(r^{ss'})\!~=~\!0$. The total energy for a mixture of $N$ particles is given by $U(r_i,...,r_N)\!~=~\!\sum_{{j>i}}u_{s(i)s(j)}(r_{ij})$.

\subsection{Initial arrested multi-gel structures}

The initial configurations are composed of $N\!~=~\!5.53\times10^{4}$
particles in a cubic simulation box of size $L$
with periodic boundary conditions. The particle volume fraction $\phi$ is estimated as the fraction of the total volume $V\!~=~\!L^3$ that is occupied by the $N$ particles $\phi\!~=~\!N\pi d^3/6V$. All gel configurations are prepared starting from a gas at $k_BT/\epsilon\!~=~\! 10$ which is quickly cooled down to $k_BT/\epsilon \!~=~\! 0.01$ and then let evolve. Afterwards, this gel configuration is further quenched by running a simulation with the dissipative viscous dynamics {as proposed in \cite{colombo2014}}:
\begin{equation}\label{eq:damp_dynamics}
m \frac{\mathrm{d^2\mathbf{r}_i}}{\mathrm{dt^2}}=-\xi \frac{\mathrm{d\mathbf{r}_i}}{\mathrm{dt}}-\nabla_{\mathbf{r}_i} U.
\end{equation}

Where $m$ is the particle mass and $\xi$ the friction coefficient. We use $m/\xi\!~=~\! 1.0 \tau^*$, with $\tau^*\!~=~\!\sqrt{md^2/\epsilon}$ the molecular dynamics time unit in our simulations, which corresponds to the overdamped limit~\cite{Nicolas2016}. The kinetic energy is drawn from the system down to a value lower than $10^{-10}$, which leads to a final arrested configuration corresponding to a local minima of the potential energy known as the inherent structure \cite{stillinger1984}. All simulations have been performed in LAMMPS\cite{plimpton1995} with a time step of $\delta t \!~=~\!5 \times 10^{-4}\tau^*$ in the initial sample preparation runs.

\subsection{Mechanical tests}

The mechanical properties of our colloidal gels are investigated through (i) small amplitude oscillatory rheology (SAOR) to characterise the linear response and (ii) continuous strain deformation for large deformations.

\subsubsection{Small Amplitude Oscillatory Rheology}

The frequency dependence response is measured by imposing an oscillatory shear strain to the system of the form $\gamma(t)\!~=~\!\gamma_0\sin(\omega t)$. Here, the new dynamics is given by
\begin{equation}
m \frac{\mathrm{d^2\mathbf{r}_i}}{\mathrm{dt^2}}=-\nabla_{\mathbf{r}_i} U- \xi \left( \frac{\mathrm{d\mathbf{r}_i}}{\mathrm{dt}}- \dot{\gamma}(t) y_i \mathbf{e_x} \right).
\end{equation}

We follow \cite{colombo2014} and compute the average state of stress of the gel $\sigma_{\alpha\beta}$ using the zero temperature virial definition~\cite{Subramaniyan2008}  ,
\begin{equation}
\sigma_{\alpha\beta}=\frac{1}{V} \sum_{i=1}^{N} \frac{\partial U}{\partial r_{i}^{\alpha}} r_{i}^{\beta}.
\end{equation}
Where $\alpha,\beta$ stand for the Cartesian components $\{x,y,z\}$. Here we have only considered the pairwise energy contribution, the kinetic term $mv_i^{\alpha} v_i^{\beta}$ as well as any contribution from the viscous drag in Eq.~\ref{eq:def_strain} are ignored as the velocities for all the particles are small for all shear rates. During each cycle, we compute the shear stress $\sigma_{xy}$ referred as $\sigma$ in the rest of the paper.

For visco-elastic solids, the shear stress can be decomposed as:

\begin{align}\label{eq:liss_harmonics}
\sigma(t) =& \gamma_0 \Big[ \left(G^{\prime}_1(\omega, \gamma_0) \sin (\omega t)  + G^{\prime \prime}_1(\omega, \gamma_0) \cos (\omega t)\right) \nonumber\\
 &+ \left(G^{\prime}_3(\omega, \gamma_0) \sin (3\omega t) + G^{\prime \prime}_3(\omega, \gamma_0) \cos (3\omega t)\right) \Big].
\end{align}
Where we have considered an expansion up to the third harmonic, with $G'(\omega, \gamma_0)$ and $G''(\omega, \gamma_0)$ the storage and loss moduli, respectively. The indexes $1$ and $3$ indicate the first and third harmonic expansion. The first term in this expression is the relevant part in the linear regime. The moduli are extracted by a simple Fourier transform of $\sigma$ and $\gamma$.

\subsubsection{Continuous strain by step deformation}

Following the same scheme as in \cite{colombo2014}, a series of incremental strain steps in simple shear geometry are applied to samples produced as previously described. For each step, the cumulative strain is increased by $\Delta\gamma$ . This is done by first applying an instantaneous affine deformation $\Gamma_{\Delta \gamma}$, which describes a simple shear in the $xy$ plane:

\begin{equation}\label{eq:def_strain}
\mathrm{\mathbf{r}}'_{i}=\Gamma_{\Delta \gamma} \mathrm{\mathbf{r}}_{i}=\begin{pmatrix}
  1 & \Delta \gamma & 0 \\
  0 & 1 & 0 \\
  0 & 0 & 1 
 \end{pmatrix}\mathrm{\mathbf{r}}_{i}.
\end{equation}
After the boundary conditions are updated, the affinely deformed configuration is relaxed by allowing the system to evolve following Eq.~\ref{eq:damp_dynamics} while keeping the global strain constant. After repeating the previous procedure for $n$ steps, the cumulative strain is $\gamma\!~=~\!n\Delta\gamma$. In our measurements, we chose a strain increment of $\Delta \gamma\!~=~\!0.01$ and a relaxation time interval $\Delta t$ such that the shear rate is $\dot{\gamma}=\Delta \gamma/ \Delta t \!~=~\!10^{-5}\tau_0^{-1}$, with $\tau_0=\xi d^2/\epsilon$ the time it takes a particle subjected to a typical force $\epsilon/d$ to move a distance equal to its size. The integration time step has been set to $\delta t\!~=~\!5 \times 10^{-3}\tau^*$ for all the mechanical tests.

\subsection{Structural {and dynamical microscopic} observables}

The structure of multi-component mixtures is characterised both prior to and during deformation by measuring the pore chord length probability distribution function $p(\ell)$. This observable is defined as the probability $p(\ell) \mathrm{d}\ell$ of finding a chord length between $\ell$ and $\ell+\mathrm{d}\ell$\cite{levitz,whittle1999} inside a pore. It is obtained by randomly sampling chord lengths on randomly chosen points. This distribution function maps the pore size, and is used here to characterise inherent structures and the effect of deformation in the pore size of the networks. During deformation we keep track of the average chord length $\langle\ell\rangle$, and use this quantity to compare the changes in different mixtures.

At the local scale, we track the dynamics of the formation and breaking of bonds. Two particles of the same species are considered bonded when the center to center distance is $r\leq1.3d$, at longer distances the energy of the bond is considered negligible. In our simulations, we keep track of the average number of bonds per particle in the mixtures, which we define as
\begin{equation}
    \langle n_b \rangle=\frac{1}{m}\sum_{{s=1}}^{m} \langle n_s \rangle.
\end{equation}
With $\langle n_s \rangle$ the average number of bonds for species $s$ and $m$ the total number of components. This quantity is calculated before and during shearing. 

The non-affine squared displacement in continuous strain deformation is measured for all the systems studied. This quantity is calculated at a step $n$ using the previous step $(n-1)$ as reference, and is defined as
\begin{equation}
    \langle \Delta^2_n \rangle=\frac{1}{N}\sum_{i=1}^{N} \left( \mathrm{\mathbf{r}}_{i,n}-\Gamma_{\Delta \gamma}\mathrm{\mathbf{r}}_{i,n-1}\right)^2.
\end{equation}
This is a per-particle quantity, and measures the variation in deformation (from the one imposed) after each step. 

\section{Results }
\subsection{Structural characterisation }

\begin{figure*}
\centering
\includegraphics[width=0.9\linewidth]{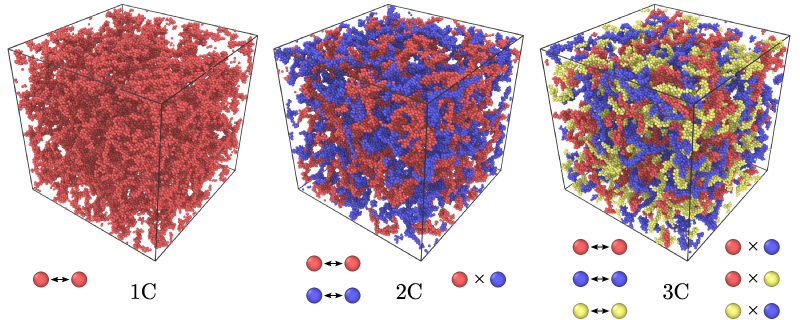} 
\caption{Snapshots of arrested phases at $\phi_{tot}\!~=~\!0.10$ for monogel, bigel and trigel. The interactions (attractive or repulsive) are also depicted for each species}
\label{fig:network}
\end{figure*}

We study three different mixtures with one (1C), two (2C) and three (3C) components. In a given mixture, all components are at equal relative concentrations $c_m\!~=~\!\phi_m/\phi_{tot}\!~=~\!1/m$, with $m$ the total number of species in each mixture. We have explored packing fractions in the range $0.10\leq \phi_{tot}\leq 0.30$. Following the quenching procedure discussed above, the three mixtures form gels characterised by a geometrically percolated network for each component. These structures are the result of an arrested phase separation. For the 1C this corresponds to a colloidal gel~\cite{foffi2005JCP, lu2008}, while the 2C configuration is originated by an arrested demixing that results in a bigel~\cite{Varrato2012, DiMichele2014}. The ternary mixture, i.e. 3C, forms the extension of a bigel, a \textit{trigel}. Typical snapshots of the final arrested configurations for the three mixtures at $\phi_{tot}\!~=~\!0.10$ are presented in Fig.~\ref{fig:network}. From the images, it is clear that bigels and trigels present intertwined networks which result from the gelation process. This by-eye observation was confirmed by a careful characterisation of the percolation of each component.

The pore chord length probability distribution function $p(\ell)$ was measured for all structures, Fig.~\ref{fig:pore} shows results for the three mixtures at three different packing fractions. {At each value of the overall packing fraction, one can see that the pore size does not changes significantly with increasing the number of components, indicating that each component is forming strands that are progressively thinner. This is confirmed by the average number of bonds per particles that is also progressively lower, with increasing the number of components, since bonds can only be formed by particles of the same species}. By comparing the packing fractions, it is evident that the distributions tend to move to lower values of $\ell$ on increasing the density. This effect is not surprising and is a consequence of the dependence of the pore size on density. More interesting is the comparison between different mixtures for the same packing fractions. At $\phi_{tot}\!~=~\!0.10$ (Fig.~\ref{fig:pore}a), the initial structures are similar for all mixtures, in agreement to what was found for bigels~\cite{Varrato2012}. Upon increasing the packing fraction, the chord length distributions for bigels and trigels start to show deviations from the monogel. At $\phi\!~=~\!0.30$ (Fig.~\ref{fig:pore}c), the highest packing fraction investigated, $p(\ell)$ shows a shift towards smaller $\ell$, which highlights the presence of smaller holes with respect to the monogel. 

The number of components at a fixed packing fraction affects locally the structure of the arms in the gels. This difference can be observed in the average number of bonds per particle $\langle n_b \rangle$, which are lower for multi-gels compared to monogels (Fig.~\ref{fig:pore}d) and lead to thinner arms. As expected, increasing the number of components exposes more surface. It is possible that the range of  attractive interactions, set in our model by the exponent values $p\!~=~\!36$ and $q\!~=~\!24$ in Eq.~\ref{eq:potentiallq}, might play an important role but this was not explored in this work.

\subsection{Gels under deformation}

\begin{figure}
\centering
\includegraphics[width=0.48\textwidth]{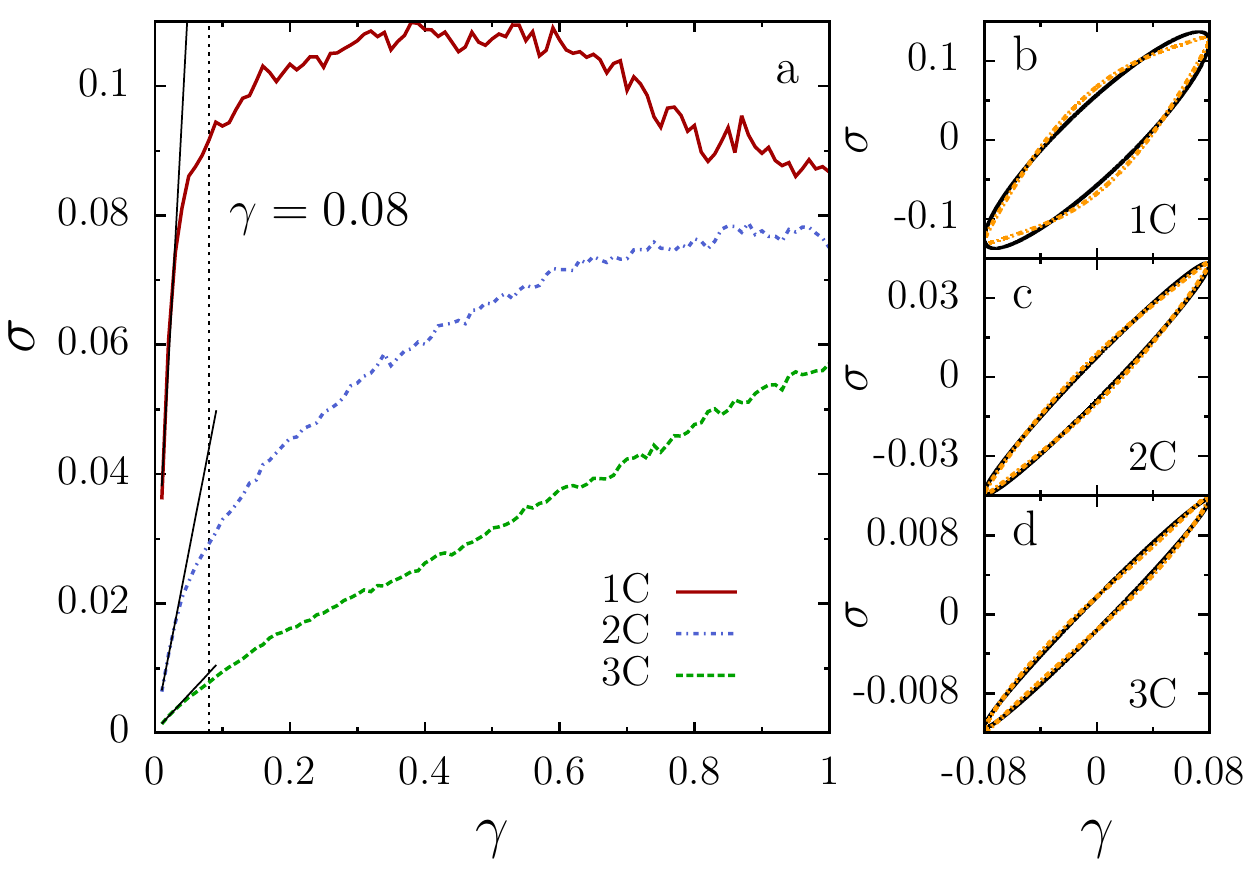}
 \caption{a) Stress vs. strain curve for monogel, bigel ($c_{2}\!~=~\!0.5$) and trigel ($c_{3}\!~=~\!0.33$). Total packing fraction $\phi_{tot}\!~=~\!0.20$, share rate $\dot{\gamma}\!~=~\!10^{-5}\tau_0^{-1}$. Solid black lines indicate the linear regimes in the three mixtures. Dashed black line marks a strain value of $\gamma\!~=~\!0.08$. b-d) Lissajous curves for a maximum strain $\gamma_0\!~=~\!0.08$ with $\omega\gamma\!~=~\!10^{-5}\tau_0^{-1}$ obtained from first and third harmonic approximations, and represented as black and orange curves respectively.}
\label{fig:step-20}
\end{figure}

The load curves (stress vs strain) for gels, bigels and trigels at $\phi_{tot}\!~=~\!0.20$ and equal relative concentrations are shown in Fig.~\ref{fig:step-20}a. We notice immediately that, in contrast with what was observed for the static quantities, the stress response displays a more complex pattern of variations  between the three systems. 

In particular, two main differences stand out. The first one is related to the location of the yielding point where the stress reaches a maximum and strong plastic effects are expected to set in. Increasing the number of components, and consequently the complexity of the structures, results in a shift of the yielding point to higher strain values. At the same time, the shear modulus at low strain, which can be interpreted as the elastic response of the material, decreases with the number of components. This unusual feature indicates that multi-gels are softer but yield at larger strains. 

The second observation pertains the linear (elastic) regime. This regime seems to extend over a higher range of $\gamma$s for more than one component, as indicated by the black solid lines in Fig.~\ref{fig:step-20}a. It appears that, the increasing complexity of the structures in multicomponent gels leads to a more stretchable material which is capable of sustaining an extended linear elastic regime. It must be stressed that the term linear used in this context refers only to the linearity of the relation between stress and strain, and not to local microscopic elastic behaviour. In other words, the response is expected always to be visco-elastic {and microscopic plastic events (such as breaking and reformation of interparticle bonds) can occur}.\\

\begin{figure}
\includegraphics[width=0.48\textwidth]{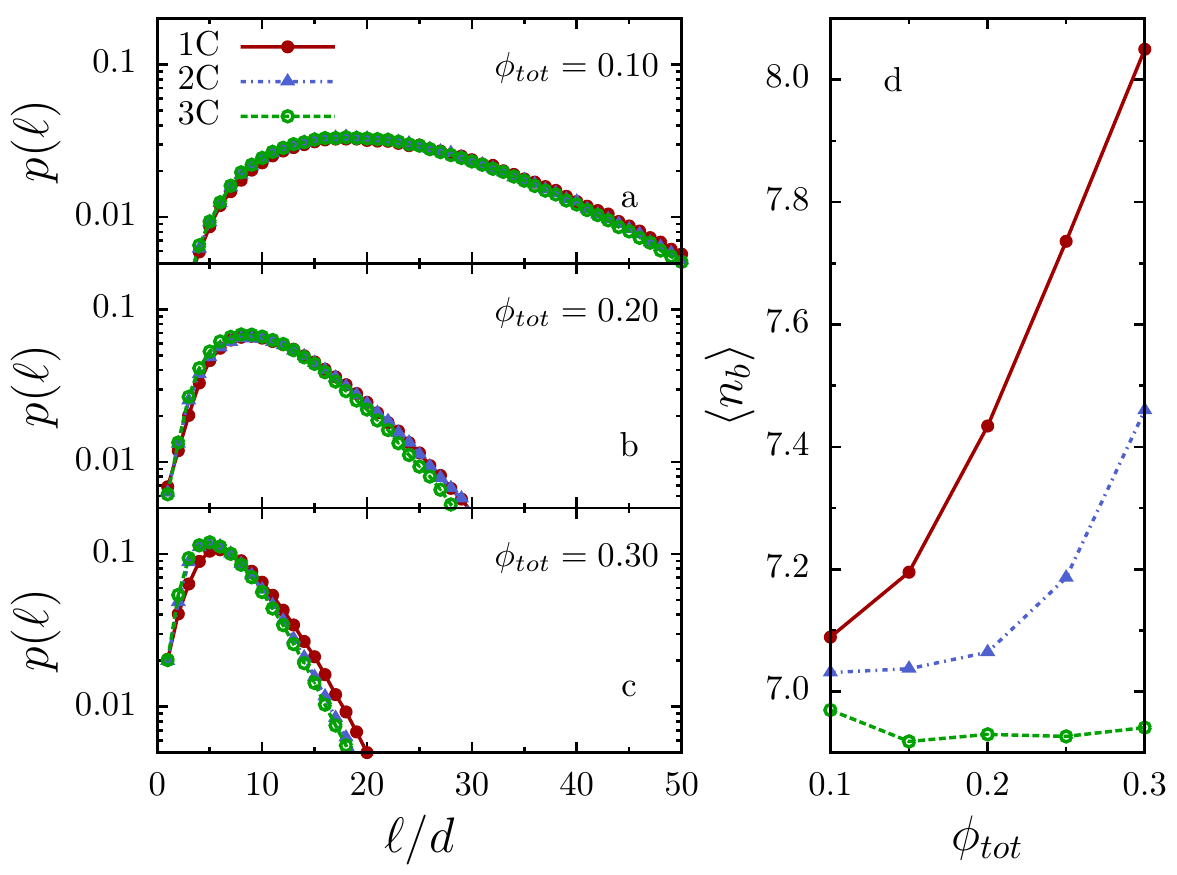}
 \caption{a-c) Effect of the number of components in the chord length distribution $p(\ell)$, for different total packing fractions. d) Average number of bonds per particle at different packing fractions. Bonds form only between same species.}
\label{fig:pore}
\end{figure}

To explore further this visco-elastic behaviour, we use small amplitude oscillatory rheology to measure the Lissajous curves (Fig.~\ref{fig:step-20}b-d) using Eq.~\ref{eq:liss_harmonics} with a maximum strain of $\gamma_0\!~=~\!0.08$ at an equivalent shear rate $\omega\gamma_0\!~=~\!10^{-5}\tau_0^{-1}$. This representation can be used to decouple the elastic contribution from the viscous one, in the stress tensor, at a fixed amplitude. This particular $\gamma_0$ value is used as a qualitative measurement to illustrate the differences between the three systems at a total strain deformation of $\gamma\!~=~\!0.08$ (dashed line in Fig.~\ref{fig:step-20}a). 

For each system, $\sigma$ is calculated using both a first harmonic approximation (black curves) and a third harmonic approximation (orange curves). The former corresponds only to the first term in  Eq.~\ref{eq:liss_harmonics}, the latter considers the full equation. In these figures, only the reconstructions up to the first and third harmonics are shown (Eq.~\ref{eq:liss_harmonics}), which are sufficient for our systems since higher order terms do not give a relevant contribution. The full simulations data are not reported as they are indistinguishable, in this strain regime, from third harmonic approximation. The monogel shows a deviation from the ideal visco-elastic regime, testified by the deviation of the Lissajous curve from a perfect ellipse, which is expected when higher order harmonics are non-negligible. On the other hand, multicomponent gels display an ideal viscoelastic behavior. We concluded that the nonlinear terms at $\gamma\!~=~\!0.08$ are more significant for the monogel than for multi-component systems. Thus, multi-component systems have an unexpected extended linear regime with respect to monogels, reinforcing the idea that more complex interdigitated gel structures have an internal stress {relaxation} mechanism that is capable of extending the linear behavior. Another interesting observation is that the area of the ellipse is shrinking as the number of components are increased, confirming the more elastic behavior of multicomponent gels. Finally, above yielding, the monogel overshoots and the stress starts to decay. This could be an indication of brittle to ductile transition, as observe in well annealed glasses\cite{Ozawa2018}, upon increasing the number of components. 

\begin{figure}
\centering
\includegraphics[width=0.42\textwidth]{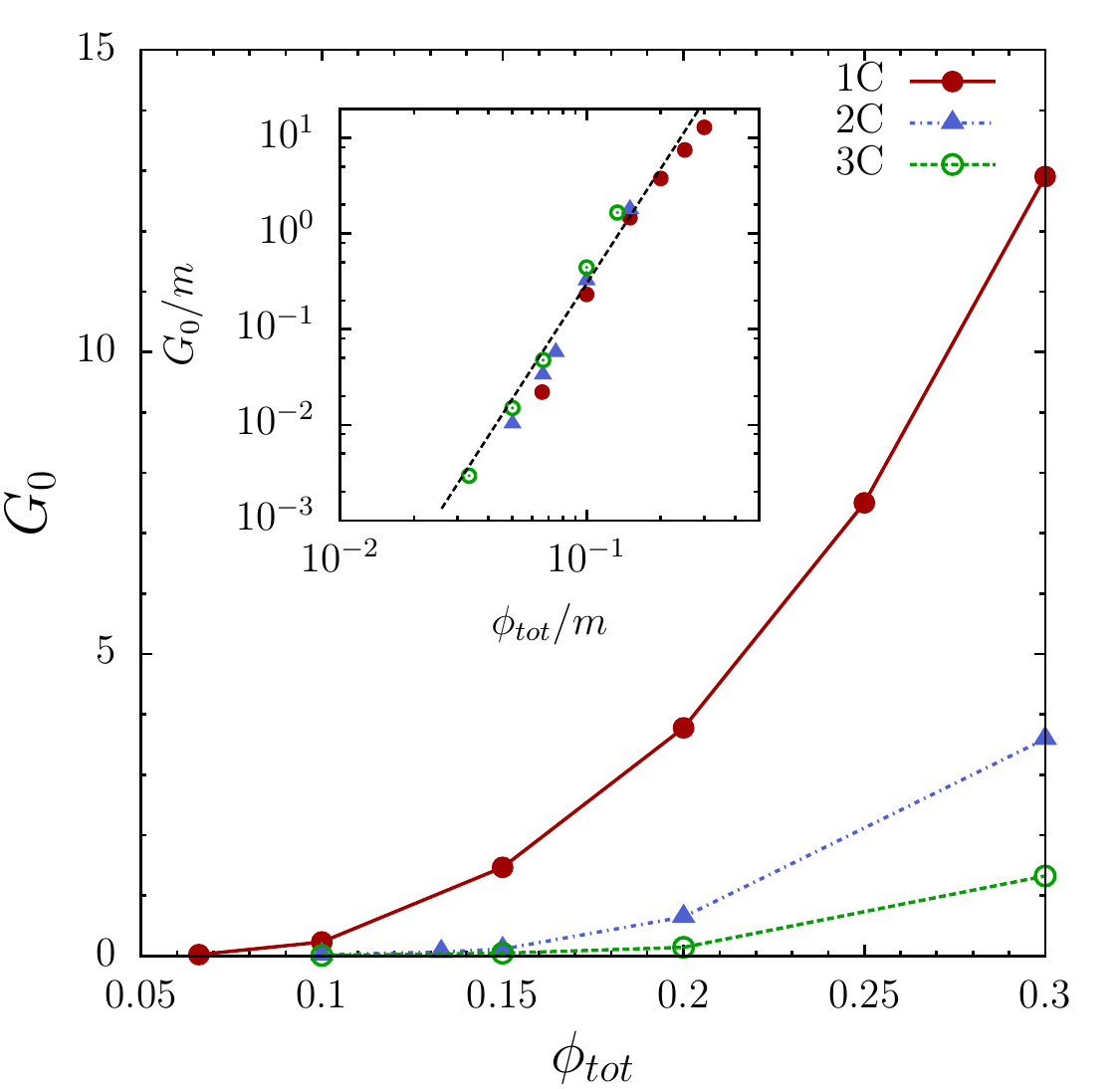}
\caption{ Low frequency elastic modulus $G_0$ as a function of total packing fraction $\phi_{tot}$. The inset shows the elastic modulus of each component $G_{o,m}$, for mixtures with $m=1,2,3$ species, at an effective packing fraction $\phi_m\!~=~\!\phi_{tot}/m$. The dashed line follows $G_{0,m} \propto \phi_m^k$ with $k=4$.}
\label{fig:g_phi}
\end{figure}

\begin{figure*}
\centering
\includegraphics[width=1\textwidth]{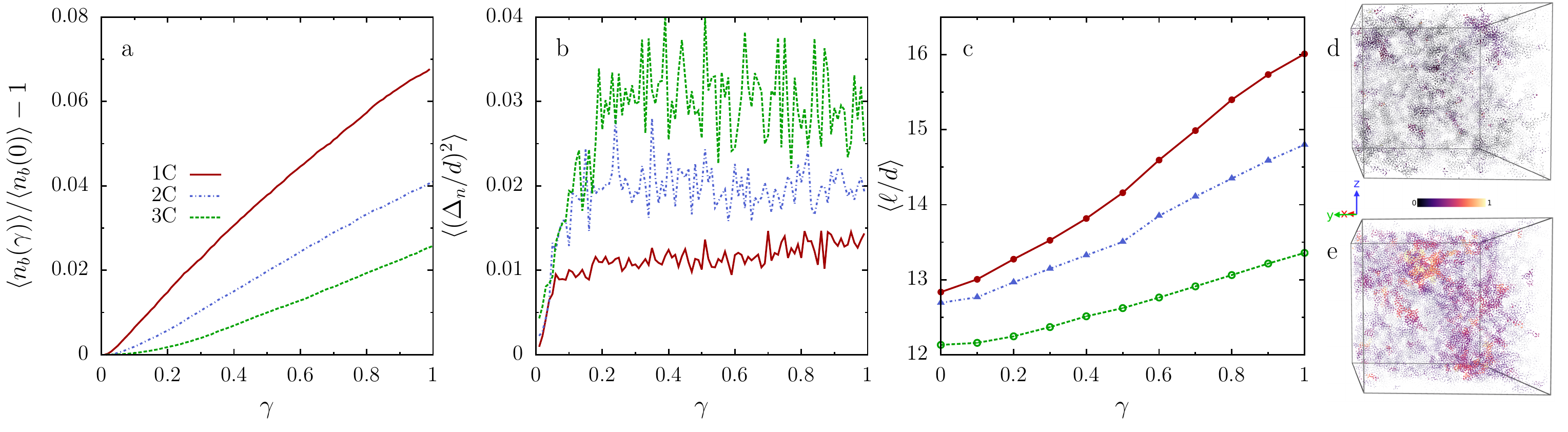}
 \caption {a) Rate of increment of the average number of bonds per particle during deformation and initial number of bonds. b) Average non-affine squared displacement vs stress. c) Average $\ell$ at different strains for all the systems. All the samples are a the same total packing fraction $\phi_{tot}\!~=~\!0.20$. Snapshots of non-affine vectors at $\gamma=0.4$ for d) monogels and e) trigels. Colour coding indicates magnitude of displacements in $d$ units (See Movies).}
\label{fig:bonds_shear}
\end{figure*}

To further explore the linear regime, we extracted the zero-frequency shear modulus $G_0$ for small deformations ($\gamma_0\!~=~\!0.01$) from the frequency dependent elastic modulus $G'(\omega)$. The results are in Fig.~\ref{fig:g_phi}, which shows $G_0$ as a function of the total packing fraction for different components. On the one hand, the fact that the gels are progressively softer, with increasing number of components, is consistent with progressively thinner strands for each component and a comparatively lower average number of bonds per particle, as indicated by the structural analysis in Fig.\ref{fig:pore}. $G_0$ grows with packing fraction for the same mixture, as expected, but decreases on increasing the number of components, which is consistent with low strain behavior in Fig.~\ref{fig:step-20}a. 
On the other hand, the dependence of $G_0$ with the volume fraction seems to change with increasing the number of components.   

A natural question that arises at this point is if we can consider each component as an independent elastic element that contributes to the overall mechanical behaviour. We propose to model our system as a set of elastic springs connected in parallel. In this approach, each component will be an elastic element and the overall spring constant will be the sum of each component's constant. Within this approximation, the low frequency elastic modulus $G_{0}$ of a multi-gel  can be written as:
\begin{equation}
    G_{0}\approx \sum_{{s}} G_{0,s}=mG_{0,m}.
    \label{G0}
\end{equation}
Where $G_{0,m}$ is the low frequency elastic modulus of an isolated component in a mixture of $m$ species. Quite naturally, we have assumed that at equal relative concentrations $G_{0,m}$ is the same for all species.
From Eq.\ref{G0}, we can see that $G_{0,m}\!~=~\!G_{0}/m$ which is a consequence of our approximation. At the same time, each component is at $m$-dependent packing fraction $\phi_m\!~=~\!\phi_{tot}/m$. If our approximation holds, the data for the different mixtures should collapse on the same master curve.

As shown in  in Fig.~\ref{fig:g_phi}, this scaling is a decent first approximation, with the difference that the elastic modulus is slightly shifted to higher values for multi-gels. This plot also indicates that the elastic modulus of each component scales as $G_{0,m} \propto \phi_m^k$, and that the exponent $k$ is roughly equal to four in the three cases. As expected, the data collapse seems to break at high densities where the microscopic structures are more different, as can be seen in the number of bonds in Fig.~\ref{fig:pore}d. In this regime, the variation in the number of bonds comes from the interaction potential, and affects the scaling in Fig.~\ref{fig:g_phi}. Nonetheless, the inset agrees with a model of springs in parallel and explains the origin of the complexity in the response to deformation for multi-gels. These results suggest that the small deformation regime can be taken into account with a simplified model that does not includes cooperative effects between the different gel strands. However, it does not explains the peculiar mechanical behaviour that is seen in Fig.\ref{fig:step-20}. This opens the question: what is the origin of the highly resilient behavior of multicomponent gels?.\\

To unravel the mechanisms at play, we now look at the microscopic configuration changes in samples during continuous deformation. The rest of the analysis focuses only on samples at a total packing fraction $\phi_{tot}=0.20$, which is chosen as a representative packing fraction for the range of values explored. First, we look at the evolution of the number of bonds. Fig.~\ref{fig:bonds_shear}a shows the rate of increment in the average number of bonds per particle, $\langle n_b(\gamma)\rangle / \langle n_b(0)\rangle-1$, at a particular strain $\gamma\!~=~\!n\Delta\gamma$ with respect to the initial configuration. From this graph, it is clear that the number of bonds increases at a higher rate for monogels than for bigels and trigels. Next, this information is complemented by analysing the spatial rearrangements during deformation with the non-affine squared displacement $\langle \Delta^2_n \rangle$. Fig.~\ref{fig:bonds_shear}b shows that the trigel displays not only the more important microscopic rearrangements for the bonding pattern but also that {particle non-affine displacements are bigger in systems of more than one component and are mainly plastic in nature (i.e. related to bond breaking or formation)}. These non-affine displacements are not only bigger but also delocalised, as can be seen in Fig.~\ref{fig:bonds_shear}d-e which show the non-affine displacement vectors for a monogel and a trigel at a strain value $\gamma=0.4$ (see Movies).

The last quantity we look at is the pore size during deformation, which characterises the evolution of the global structure. Fig.~\ref{fig:bonds_shear}c shows the average chord length $\langle \ell \rangle$ for the three mixtures during deformation, with a smaller increment for multi-gels which indicates less macroscopic structure changes than in the monogel. Fig.~\ref{fig:bonds_shear}c also shows that in trigels $\langle \ell \rangle$ deviates more slowly from the initial values than in monogels, indicating that trigels keep their macroscopic structure for longer.

Fig.~\ref{fig:step-20} and \ref{fig:bonds_shear} convey an unexpected picture. On one side, systems with a larger number of components present an extended linear regime (Fig.~\ref{fig:step-20}) but, at the same time, they display  more dramatic microscopic rearrangements (Fig.~\ref{fig:bonds_shear}b). This hints to a \textit{self healing} mechanism which {helps} the gels {restructure} via large non-affine displacements, due to the steric hindrance between non interactive components, maintaining their original {overall load-bearing} structure for longer.

Such results provide a more clear picture of the microscopic and macroscopic behaviour of multi-component gels. The presence of multiple components delays the permanent damage of the gels when deformed, and allows these materials to maintain a similar pore structure during deformation for bigger strains than a monogel. This toughening mechanism is presumably due to the activation of low frequency modes through the excitation of softer regions that accommodate the strain by deforming the gel along the direction that cost less energy through higher non-affinity.

\section{Conclusions}
In this work we have investigated the link between the structure and the mechanical properties of a new class of gels made by intertwined (and mutually repulsive) multi-component networks. At low density, the structural analysis shows similar pore size distribution widely spread over several length scales, that tend to homogenise with increasing the overall density. Multi-component gels form thinner arms that { lead} to a decrease in the shear modulus as well as an extended linear response under deformation. These soft structures {provide} the possibility to self-reorganise and heal the gels, accommodating larger deformations. The physical mechanism involves large non-affine deformations originated from steric interactions that prevent a local compactification of the structure. Our work shows that the combination of this collective reorganisation of the arrested structures and the interspecies interactions can lead to new mechanical properties. \\
The model presented here is clearly simplified but the mechanism for the unusual mechanical behaviour is a result of the interplay between the interpenetrating networks, a situation that is encountered in a lot of soft matter systems. In particular, the preservation of the network structure at larger deformations can be of great importance for a lot of applications in which one looks for soft materials that are able to preserve their connectivity such as  soft robotics\cite{Majidi2018}, dielectric elastomers\cite{OHalloran2008} and  stretchable electronics\cite{Rogers2010}.\\
Similar forms of gels have been been realised recently using different approaches, but we want to stress an important point. One of the main features of our model is the fact the individual gels  that compose the multi-gels are identical. This is a substantially different situation to double polymeric network gels where properties such us self healing and toughness are encoded independently in the two networks\cite{Gong2010}. In our case, the unusual mechanical properties are solely a consequence of the inter-gels interactions. In this sense, our work tries to shed a light on the physical behavior of multi-gels as originated solely by the cooperative effects between the individual components. In the future, {we would like to investigate the role of possible variations of} the types of components. It would be interesting, for example, to see what happens when the bond energies of one of the species are varied and {different interactions between the components are possible}. Another aspect, that we have only partially addressed, is the effect of varying the protocol to gelation. As a matter of fact, gels are out-of-equilibrium solids and their thermal history is highly coupled to their structure. Finally, it would also be interesting to look more in detail into fatigue tests. It has been shown that glasses, under periodical deformation, are able to go under annealing and hardening\cite{Leishangthem2017}. Gels, with their heterogeneous structure, { are intriguing for us in this respect}.

\section*{Acknowledgements}
We thank S. Sastry for constructive comments on this work.

\end{document}